\def\rx16{1RXS~J161008.0+035222\ }
\def\rrx16{1RXS~J161008.0+035222}
\def\rxx23{1RXS~J231603.9$-$052713\ }
\def\rrxx23{1RXS~J231603.9$-$052713}
\documentclass[useAMS,usegraphicx,usenatbib]{mn2e}
\usepackage{times}

\title[Optical and infrared observations of \rx16 and \rxx23]
{Optical polarimetry and infrared photometry of two AM~Her binaries: \rx16 and
\rxx23
\thanks{Based on observations made at the Observat\'orio do Pico dos Dias,
Brazil,
operated by the Laborat\'orio Nacional de Astrof\'\i sica.}}
\author[C.~V. Rodrigues et al.]
{C.~V. Rodrigues,$^{1}$\thanks{E-mail: claudiavr@das.inpe.br}
F.~J. Jablonski,$^{1}$
F. D'Amico,$^{1}$
D. Cieslinski,$^{1}$
J.~E. Steiner,$^{2}$
\newauthor
M.~P. Diaz,$^{2}$
and G.~R. Hickel$^{3}$\\
$^{1}$ Instituto Nacional de Pesquisas Espaciais/MCT --
Av. dos Astronautas, 1758 -- 12227-010 - S\~ao Jos\'e dos
Campos - SP -- Brazil\\
$^{2}$ Instituto de Astronomia, Geof\'\i sica e Ci\^encias Atmosf\'ericas/USP --
Rua do Mat\~ao, 1226 -- 05508-900 - S\~ao Paulo - SP -- Brazil\\
$^{3}$ IP\&D - Universidade do Vale do Para\'\i ba -- Av. Shishima Hifumi, 2911
--
12244-000 - S\~ao Jos\'e dos Campos - SP -- Brazil}

\begin{document}



\maketitle

\label{firstpage}

\begin{abstract}
We present the first optical circular and linear polarization 
measurements of two polar candidates from {\it ROSAT}: \rx16 and \rrxx23.
We also present {\it H} band near-infrared photometry of the last object.
The presence of  strong circular polarization
confirms them as AM Her systems. \rxx23 was observed in two
different brightness states. The orbital phase dependence of the
flux and polarization of \rx16 is reasonably fitted with a simple model 
in which the binary is observed at a small inclination and the magnetic field
axis is almost parallel to the white-dwarf rotation axis resulting in the
accretion column axis being
seen from top during the whole orbital revolution. An alternative
geometry with intermediate inclination can fit the observed flux and circular
polarization. However, in this case, the model produces a linear polarization 
peak which is not corroborated by the data.  The estimated magnetic
field is in the 10 to 20~MG range.
The circular polarization of \rxx23 is complex and highly variable.
The light-curves of that object have been fitted using a model which includes
the white-dwarf, a heated secondary and a point-like accretion region. 
The secondary
emission contributes significantly even  in optical wavelengths.
This model also reproduces the main features of the
optical polarization of \rrxx23. We estimate the main parameters
of the binary, of the accretion region and the distance to the system.
An improved description of this system should include an extended
and inhomogeneous accreting region as well as non-radial accretion.

\end{abstract}


\section{INTRODUCTION}

Cataclysmic variables (CVs) are short period binaries consisting of a 
white-dwarf
(primary) and a late-type main-sequence star. The secondary star fills its Roche
lobe,
losing material to the primary by the inner Lagrangian point, $L_1$. Due to its
angular momentum and viscous processes, 
this material forms an accretion disc around the white-dwarf.
Polars, also called AM Her systems after their prototype, are CVs in which the
primary has a magnetic
field in the 10 to 200 MG range. In these systems the material
from $L_1$ follows a
ballistic trajectory on the orbital plane (horizontal stream) up to the magnetic
coupling region. From this region on the gas flow effectively traces the geometry
of the magnetic field, forming an accretion stream. 
Unlike non-magnetic CVs,
no disc is formed. Another consequence of the
strong magnetic field is the synchronization of the white-dwarf rotation with
the
orbital revolution. The variety of physical processes occurring and 
many open issues regarding the details of the emission components 
in these systems make them very interesting targets for
astrophysical research.
Reviews on polars can be found in Cropper (1990) and Warner (1995).  

In the accretion region, near the white-dwarf surface, the material is fully
ionized producing highly polarized cyclotron emission due to the presence of
a strong magnetic field. As a large fraction of
the optical flux in polars comes from this region, it shows large
polarization, both linear and circular. Consequently, polarimetry
is a fundamental tool to classify an object as an AM Her system. 

Stokes parameters are also powerful to diagnose the state of the
accreting region because they depend
strongly on the angle by which that region is observed as well
as on its physical properties. 
If the accretion region is small, the orbital behavior of the linear
polarization position angle can be described
by a simple equation that depends on the inclination
of the system, $i$, and the colatitude of the
axis of the magnetic field, $\beta$. 
Besides the quantities mentioned above, the phase interval during which 
no cyclotron emission is observed also constrains $i$ and $\beta$.
Estimates of the value of the magnetic field as well as of some other plasma
properties can also be obtained through the modelling of flux and polarization
variations with orbital phase (see Wickramasinghe \& Meggitt 1985 
and references therein on accreting column cyclotron models). 

The fall of the column material onto the white-dwarf surface produces hard
X-ray emission mainly through bremsstrahlung which is reprocessed
as thermal soft X-ray emission. Consequently, {\it ROSAT} satellite has 
discovered many polars, contributing significantly to the number of 
known systems (see Schwope et al. 2000a, for instance).

In this work, we present new data for {\it ROSAT} polar candidates:
$R_C$ band polarimetry of \rx16 and $R_C$ and $I_C$ band polarimetry and {\it H}
band
photometry of \rrxx23. Some modelling of the objects is also presented. 
Preliminary results of these observations were presented in Rodrigues et al.
(2005).

\section{OBSERVATIONS AND DATA REDUCTION}

In the following sections, we describe the acquisition and reduction 
of the polarimetric and infrared data. A summary of all observations is
presented in Table \ref{tab-obs}.

\subsection{Optical polarimetry}

The observations have been performed with the 1.6-m Perkin-Elmer
telescope at the {\it Observat\'orio do Pico dos Dias} (OPD),
Brazil, operated by
the {\it Laborat\'orio Nacional de Astrof\'\i sica} (LNA), Brazil. We have used
a
CCD camera modified to contain a polarimetric module described in
Ma\-ga\-lh\~aes et
al. (1996). 
The instrument consists of a fixed analyzer (calcite prism),
a $\lambda$/4 retarder waveplate and a filter wheel. The retarder plate
is rotated in 22\fdg5 steps. A complete measurement
consists of eight images in consecutive retarder
orientations.
The calcite block separates the extraordinary and
ordinary beams by 12\arcsec. This technique eliminates any sky
polarization (Piirola 1973; Ma\-ga\-lh\~aes et al.  1996). The $\lambda$/4
retarder allows us to measure circular and linear polarization
simultaneously.
The CCD arrays were SITe back-illuminated devices, with $1024 \times 1024$
pixels.

\begin{table*}
 \centering
 \begin{minipage}{165mm}
  \caption{Log of observations}
  \label{tab-obs}
  \begin{tabular}{@{}cccccccc@{}}
  \hline
Object  & Date &  Telescope & Instrument & Filter & Retarder & Integration time
& Time span\\
        &      &            &            &        &          &  [s] & [h] \\
\hline
1RXS J161008.0+035222 & 2003 Apr 22 & 1.6-m & CCD camera & $R_C$ & $\lambda$/4 &
90 & 2.4 \\
                      & 2003 Apr 23 & 1.6-m & CCD camera &  $R_C$ & $\lambda$/4
&
80 & 3.6 \\
                      & 2003 Apr 26 & 1.6-m & CCD camera &  $R_C$ & $\lambda$/2
&
60 & 4.2 \\
\\
1RXS J231603.9$-$052713 		& 2003 Jul 04 & 0.6-m & CamIV & $H$ & - & 60 &
2.8 \\
			& 2003 Jul 05 & 0.6-m & CamIV & $H$ & - & 60 &
2.1 \\
			& 2003 Jul 06 & 0.6-m & CamIV & $H$ & - & 60 &
2.5 \\
\\
& 2003 Sept 24 & 1.6-m & CCD camera  & $R_C$ & $\lambda$/4 & 80 & 3.3\\
& 2003 Sept 28 & 1.6-m & CCD camera  & $I_C$ & $\lambda$/4 & 90 & 1.3\\
\\
			& 2004 Jun 22 & 1.6-m & CamIV & $H$ & - & 30 &
1.6 \\
			& 2004 Jun 23 & 1.6-m & CamIV & $H$ & - & 30 &
1.7 \\
\\
			& 2004 Jul 28 & 1.6-m & CamIV & $H$ & - & 30 &
2.1 \\
\\
                        & 2004 Oct 06 & 1.6-m & CCD camera  & $R_C$ &
$\lambda$/4
& 90 & 2.1\\
                        & 2004 Oct 07 & 1.6-m & CCD camera  & $R_C$ &
$\lambda$/4
& 90 & 5.2\\
                        & 2004 Oct 08 & 1.6-m & CCD camera  & $R_C$ &
$\lambda$/4
& 90 & 6.0\\
                        & 2004 Oct 09 & 1.6-m & CCD camera  & $I_C$ &
$\lambda$/4
& 90 & 4.3\\
\hline
\end{tabular}
\end{minipage}
\end{table*}

The images have been reduced following standard procedures using
{\sc IRAF}\footnote{{\sc IRAF} is distributed by
National Optical Astronomy Observatories, which is operated by the Association
of Universities for Research in Astronomy, Inc., under contract with the
National Science Foundation.}. The extracted counts for the individual
beams were used to calculate the polarization using the method described in
Ma\-ga\-lh\~aes,
Benedetti \& Roland (1984) and Rodrigues, Cieslinski \& Steiner (1998). 
The polarimetric reduction was greatly facilitated by the
use of the package {\sc PCCDPACK} (Pereyra 2000). Photometry can be done
combining
the counts in the two beams. We performed differential
photometry of our target using brighter and fainter comparison stars in the field.

Our main goal was to measure the circular polarization with an
error of the order of 1\%  and a time resolution enough
to appropriately sample the orbital variability. Therefore, we chose
exposure times, $t_{int}$, in the 60--90~s range. In
this way, a polarization measurement spans 8~$\times$~$t_{int}$
plus the dead time. We would
usually group the images in sequences of 8 images with no overlap
(images 1 through 8, images 9 through 16 and so on). However, in order
to improve the temporal resolution, we have grouped the images with
overlap (images 1 through 8, images 2 through 9 and so on).  In this
way, we have a time interval between two points of typically 130~s.
However, it should be noted that the actual time resolution is worse than 
that, by a factor of  $\approx 6$, since the points are not independent. 

Each night we observed polarimetric standard stars (Serkowski, Mathewson \&
Ford 1975; Bastien et al. 1988; Turnshek et al. 1990) in order to calibrate
the system and estimate the instrumental
polarization. The measured values of the unpolarized standard stars were
consistent with zero within the errors: consequently no instrumental correction
was applied. Linear polarized and unpolarized standard stars also show
insignificant values of circular polarization. Our measurements are
available at http://www.das.inpe.br/$\sim$claudia.
Data using a Glan filter were also collected to estimate the 
polarization measurement efficiency
of the instrument. They indicate that no instrumental correction is needed.
Unfortunately, we could not calibrate our measurements in order to know
the correct sign of the circular polarization. However, we can distinguish
polarization of different signs. The signs in the figures are the instrumental
ones.

\subsection{Infrared photometry}

The data were obtained with the 0.6-m Boller \& Chivens and 1.6-m
Perkin-Elmer telescopes at OPD (see Table \ref{tab-obs} for details). 
The instrument used to collect the data is the CamIV imager, based on a
Hawaii 1k $\times$ 1k Rockwell Int. array. The image scale is 0.48 arcsec/pix in
the 0.6-m telescope, with a field-of-view (FOV) of 8\arcmin $\times$ 8\arcmin, 
and 0.24 arcsec/pix in the 1.6-m telescope, with a FOV of 4\arcmin $\times$
4\arcmin. All observations were done in
the {\it H} band. Our Hawaii detector is operated with a gain of
$4.5$~e$^-$/ADU
and has a rms read-out noise of 14.5~e$^-$. 

The telescope was offset between
exposures by 10--20 arcsec to provide estimates of the
sky background level by taking the median values of the
images at each point. The resulting time
series is not equally spaced in time, but has a time resolution
that is quite sufficient for exploring the presence of the orbital
modulation expected either from ellipsoidal variations 
or aspect variations from an asymmetrically heated
secondary star.

The data reduction was
carried out under {\sc IRAF} and consists of the following steps:

\begin{description}

\item {\sc Linearization of the images.} A simple quadratic model with
non-linear coefficient of $3.6 \times 10^{-6}$ was applied to the raw
counts in all
pixels in all images. This value has been monitored since the beginning
of the use of the detector on 2000 July, and has not changed;

\item {\sc Master flat-field image construction (for each night).} For this,
we obtained two sequences of 100 exposures of 2~s integration
time; one with the dome flat-field lamp on and the other with the
lamp off. The difference between the median-combined images
(on-off) was normalized to unity and is the master flat-field
image. Its rms measured in a \verb"[100:200,100:200]" 
section is $\sim$5\% in all nights. The quality of the dome
on-off flat-field image is comparable to what is obtained from 
taking the median values of 
all program images but it has the advantage of having being
obtained from exactly the same number of images each night;

\item {\sc Bad Pixel Map (BPM) construction.} This is done by examining
the histogram of the normalized master flat-field image. Pixels
with response $\leq 0.8$ or $\geq 1.2$ are flagged as bad. Also,
column 513 of the array is flagged for later interpolation with
the task \verb"FIXPIX".
\end{description}

The extraction of the differential light-curves is as follows. Take
for example the $j^{th}$ image in the sequence of images obtained
in one night. A ``local" sky is built from the median of $N_{sky}$
images around the $j^{th}$ image in the series. The $j^{th}$ image itself
is included in this sky estimate. We verified that $N_{sky}~=~7$
gives good results. The sky image is subtracted from the $N_{sky}$
images centred in image $j$ and each one is divided by the master
flat-field image and masked with the BPM. With the aid of the
position of the reference star in each image, all $N_{sky}$ images
are registered to the position of the $j^{th}$ image, combined, and the
photometry of the resulting combined image
is performed with \verb"PHOT/IRAF".  The Heliocentric Julian
day, reference star raw magnitude and magnitude differences of the
target and comparison stars are output to a file. The image index
is incremented to $j+1$ and the procedure is repeated. The
processing of a whole night of data does not need the user's
intervention. Both the beginning and the end of the sequence of
the images fail to have the $j^{th}$ image centred on the sky
estimate. We simply use the first (or last) $N_{sky}$ images until
(after) image $N_{sky}/2+1$ (or $j_{last}-N_{sky}/2$) is reached.

The combination of images improves the quality of the extraction
at the expense of a poorer time-resolution in the light-curve. In
our case, the time resolution is about 8 minutes in 2003 and
5 minutes in 2004. Both are good
enough to sample phenomena that are expected to have typically
$\ga$~100~minutes of time-scale.


\section{1RXS J161008.0+035222}
\label{result-rx16}

\rx16 is a {\it ROSAT} source presented in Schwope et al. (2000a), who have
already
quoted it as a possible AM Her system. Its optical
identification and spectral classification as a potential magnetic CV were made 
by Jiang et al. (2000) and
Schwope et al. (2002). These spectra show evidence of a M5 V secondary star.
In spite of having an estimate for its orbital period of 0.130\,34 d (Schwope et
al. 2002), no light-curve has been published.

\subsection{Results from optical photometry and polarimetry}

\begin{figure}
\includegraphics[width=84mm]{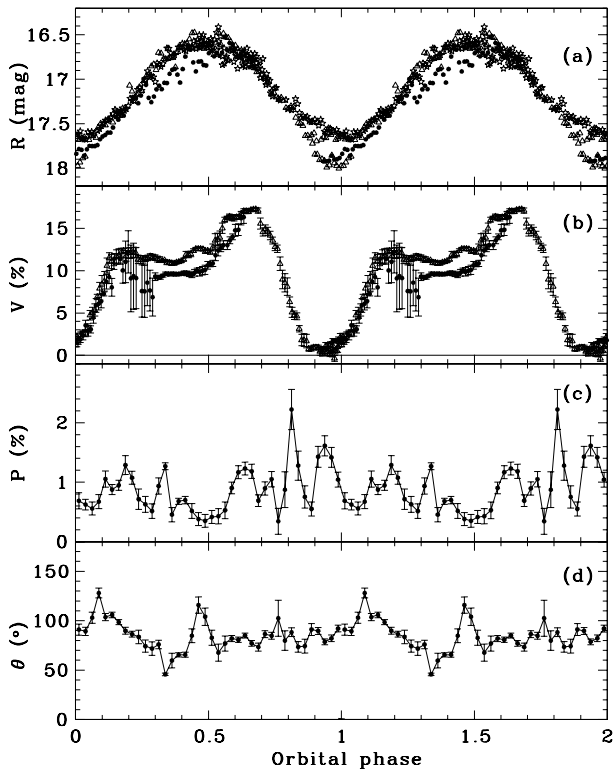} 
\caption{Observations of \rx16 in the $R_C$ band done in 2003~April.
(a) Photometry, (b) circular and (c) linear polarization, and
(d) polarization position angle. The different symbols in panels (a) and
(b) represent distinct days of observations: full dots stand for April 22,
open triangles stand for April 23 and open stars stand for
April 26. In panels (c) and (d) the data from all days were binned
in 40 phase intervals. These data were plotted assuming the ephemeris
of Eq. \ref{eq-rx16}. The circular polarization sign is the instrumental one.}
\label{rx1610_data}
\end{figure}

Figure \ref{rx1610_data}.a shows the calibrated light-curve obtained from the
differential
photometry of \rx16 with respect to the star USNO~B1.0~0938$-$0260775
$R_C$~=~17.1~mag).\footnote{The USNO accuracy is approximately 0.3~mag 
(Monet et al. 2003).} The typical uncertainty of the 
points is 0.03 mag. There are three estimates to the $R_C$ magnitude of
\rx16 in the USNO catalogues: $R_C$~=~15.8~mag (USNO~A2.0~0900$-$8458221), and
{\it R1}~=~16.3~mag and {\it R2}~=~18.99~mag (USNO~B1.0~0938$-$0260772). 
The comparison of our measurements with the USNO magnitudes suggests
that the object was in the high-state during our measurements. This
is consistent with the finding of Schwope et al. (2002) that
the light-curve shows a quasi-sinusoidal shape 
associated to the object's high mass-transfer state.

Our data can be used to constrain the ephemeris of \rrx16. 
The published period of 0.130\,34~d (Schwope et al. 2002) 
is inconsistent with our data: using this value the minima 
are out-of-phase.  Therefore we have performed
a period analysis that provides the following ephemeris for the time
of minimum flux in the $R_C$ band:

\begin{equation}
T_{min} (HJD) = 2\,452\,752.756\,7\;(70) + 0.132\,32\;(4)\;E\; .
\label{eq-rx16}
\end{equation}

This was calculated using a {\sc DFT} algorithm as in Deeming (1975).
The errors were estimated by a chi-square fitting of a sinusoidal curve to
the observational data.
Other methods, {\sc CLEAN} (Roberts, Lehar \& Dreher 1987) and the {\sc
PDM} in {\sc IRAF}, give periods of 0.132\,29\,d and 0.132\,41\,d, respectively.

Figure \ref{rx1610_data} also shows the circular (panel b) and linear
polarization (panel c) as well as its position angle (panel d).
This object shows high values of circular polarization 
clearly modulated with the orbital period. This indicates that a cyclotron
component is a significant part of the total flux of the system in the
$R_C$ band. Therefore, \rx16 is definitely a polar.

The photometry and circular polarimetry are consistent with a one-pole
accretion geometry in which the accretion region is never out of sight, 
so no occultation by the white-dwarf occurs.

The linear polarization is small along all orbital phases and
shows no conspicuous peak -
the point in the binned curve at $\sim \Phi_{orb}~=~0.8$ is not
significantly above the statistical fluctuations. We expect
peaks in the linear polarization when the column
is seen edge-on as it rotates in or out of view
over the limb of the white dwarf primary. Hence
the absence of linear
polarization peaks is also consistent with an one-pole 
geometry.

The foreground polarization produced by the interstellar dust
 in the line
of sight to \rx16 was estimated using the data of 2003 Apr 26
collected with a $\lambda/2$ retarder.
The weighted average of the linear polarization of 52 stars in the field is
0.442 $\pm$ 0.065\% at 82.0  $\pm$ 4.2$\degr$:
the quoted errors are the standard deviation
of one measurement.
The observed position angle of \rx16 is approximately constant.
This indicates that the measured linear polarization is small but not exactly 
equal to zero - in which case the position angle should have been randomly
distributed between 0 and 180$\degr$.  
If we average all the phases of \rrx16, we obtain 0.667  $\pm$ 0.017\% at
78.7 $\pm$ 0.7$\degr$. The similarity between the position angles 
indicates that the observed linear polarization of \rx16
could have an interstellar origin. We will return to this discussion in
the next section.

\subsection{Modelling the flux and polarization}

We can go a step further trying to fit a very simple model to \rrx16.
We consider two components to the system emission: one has a
constant and unpolarized flux; the second is cyclotron plus free-free in
origin, and represents
a point-like accretion region in the white-dwarf surface. 
The fluxes and polarizations (linear and circular) of the
latter component are those from  Wickramasinghe \& Meggitt 
(1985, hereafter WM85). 
Recently, improved calculations of cyclotron emission in
the accretion columns of polars have been presented (Potter et al.
2002; Potter et al. 2004). However, they are not available in
tabular format, so we could not use them in our calculations.
To compare the model with the observations, the
maximum observed flux was normalized to unity. No normalization is needed to
compare the polarimetric data with the model.
A multiplicative factor for each component (WM85 and unpolarized) is
calculated forcing the total model flux at the time of maximum light to be 1.
The models have been calculated for a wavelength of 6400~\AA.

\begin{figure}
\includegraphics[width=84mm]{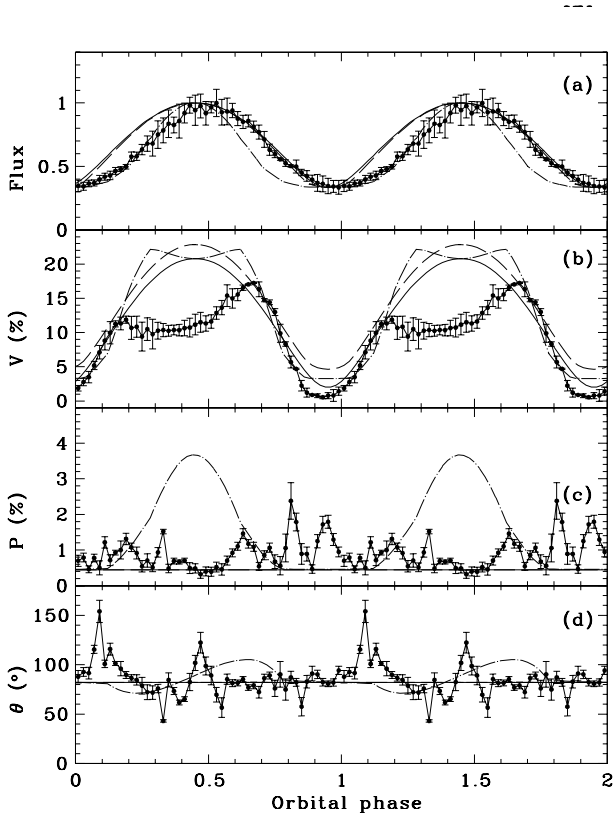} 
\caption{Simple models fit to \rrx16. The continuous and dot-dashed lines represent the 
point-like cyclotron emission region models i and ii, 
respectively (see text). 
A model with an extended region of $1\degr$ x $10\degr$ is shown by the dashed line 
(see text). The points represent all the available data binned in 50 phase intervals.}
\label{rx1610_modelo}
\end{figure}

We produced a grid of models varying the magnetic field, the
inclination of the system, $i$, and the colatitude
of the magnetic field axis, $\beta$, for the eleven WM85 tables.
We then fitted the models to the data. Each fit produces three
$\chi^2$ values. Each of them
compares theoretical predictions with the observed intensity, 
linear and circular polarization data. The models which
presented  small values of the three $\chi^2$ were inspected visually. 
Two regions in the parameters space produce models able to fit the observed
flux and circular polarization. 
Two illustrative models of each region are shown superimposed on the data in Fig.
\ref{rx1610_modelo} (panels a, b and c): model i is represented by
a full line; and model ii, by a dot-dashed line. 
Sample model parameters are: 

\begin{enumerate}

\item $i~=~2\degr$; $\beta~=~1\degr$; 
$|B|$~=~10~MG; T~=~40~keV; and $\Lambda$~=~$10^5$; 

\item $i~=~58\degr$; $\beta~=~22\degr$; $|B|$~=~8.5~MG; T~=~10~keV;
 and $\Lambda$~=~$10^7$. 

\end{enumerate}

\noindent The parameter, $\Lambda$, is the optical
depth parameter and depends on the path length, electron
number density and magnetic field (WM85). 
We added the foreground polarization obtained in section 3.1  
to the linear polarization of the models. 
The proportion of unpolarized to WM85 components in models
i and ii are 0.26/0.74 and 0.22/0.78, respectively.
Using the same geometry of model i, reasonable fits are also obtained
with $|B|$~=~16~MG, T~=~40~keV and $\Lambda$~=~$10^3$ or $|B|$~=~20~MG 
and the 20~keV shock front model of WM85.
The main difference between the two models is the presence of a linear
polarization peak around phase 0.5, which seems not to be present
in the data.  If the above solutions are valid, we can say that the
magnetic field in \rx16 is in the range 10--20~MG. 
The lack of a proper treatment of absorption/scattering
in the accretion flow by WM85 can modify substantially the results
 in the phase interval 0.7 -- 1.3. 

In a radial and point-like model for the polarized component,
the position angle of the
linear polarization is a function of two
geometric parameters, $i$ and $\beta$ (Stockman
1977; Brainerd \& Lamb 1985). This happens because
the polarization position angle is parallel or perpendicular to the 
projected angle of the magnetic field, which is parallel to the accretion
column.  Fig. \ref{rx1610_modelo} (panels c and d) shows the results for
the linear polarization which was considered as the sum of the
model and a foreground component represented by the field  stars
average. As the linear polarization of model i is approximately
zero, the resultant polarization is that of the field stars (a constant
module and angle with phase). For model ii, a peak reaching more
than 3\% is produced. As this peak is not observed in the data,
we favour the interpretation that the accretion region is seen
approximately from top in any orbital phase. However,
due to the simple assumptions of our models,
the exact values of the inclination, $i$, and the magnetic 
colatitude, $\beta$, of model i should be interpreted with care.
A spectroscopy study of this object can be helpful in constraining
the viewing angle to the column.


We forced the time of inferior conjunction of the accretion column
to be at phase 0.95. This was done in order
to better fit the circular polarization. A slightly
different choice (values nearer 1.0) could improve the flux fit instead,
shifting the models to later phases (see Fig. \ref{rx1610_modelo}).
This small disagreement in phasing between polarization and flux data
could be due to the variation of the phase of photometric minimum 
from night to night which is illustrated in Fig. \ref{rx1610_data}. 

To examine the influence of an extended emission region, we have
done some additional modelling.
We have used a grid of 25 points ($5 \times 5$) to represent
a rectangular region of arbitrary sides. All the points have
the same emission properties. Considering
$i~=~2\degr$ and $\beta~=~1\degr$ (as in model i)
and an emission region extending $1\degr$ in
latitude and $10\degr$ in longitude, 
the model produces results very similar to what the
point-like model does, but   
the minimum of the circular polarization is  
shallower than the observed (Fig. \ref{rx1610_modelo}, 
dashed line).  

The harmonic number of the cyclotronic emission can be calculated using
the adopted geometric model and the observed polarizations. Specifically,
it is proportional to the ratio between circular and linear
polarizations. The harmonic number is also inversely proportional to
the magnetic field strength, which can then be calculated from the
polarization ratio (see Wickramasinghe \& Meggitt 1982 for the theory and
Cropper, Menzies \&
Tapia 1986 for an example of application). We obtained for \rx16 data harmonic
numbers
around $10^{-3}$, which is obviously incorrect. 
A more consistent result would be obtained if the linear polarization
(produced by the cyclotron emission) was smaller than the observed. This could
be the case
if the observed linear polarization was not produced in the accretion
region but had a different origin, as in the interstellar medium, as 
previously discussed.

In the above discussion we considered that  flux modulation is totally
originated from the cyclotron emission. The results can be modified
if there is another source of modulated emission in {\it R} band in
the system.

It should be noted that a more realistic model for the accretion region must
include an inhomogeneous and extended structure. Such refinements
could produce irregular, asymmetric curves as observed 
in polars in general, and also in \rrx16.  

\section{1RXS J231603.9$-$052713}

\rxx23 is also a polar candidate from {\it ROSAT} (Beuermann \& Thomas 1993;
Thomas et al.
1998; Schwope et al. 2000a). No optical study has been published on this 
object so far. The orbital period quoted in Downes et al. (2001) catalogue is 
$0.145\,45\;$d (Thomas H.-C., 2004, private communication).

\subsection {Orbital period determination}

The distribution of the observing runs in 2003 was such that
one cannot unambiguously count the number of orbital cycles to the
2004 season. A provisional value for the period of the orbital modulation
from the combined optical and infrared data in 2003 is $0.146\,13 \, \pm \,
0.000\,14$~d.
This is marginally consistent with the value in Downes et al. (2001) and
a good starting point to analyze the 2004 data. As shown in Fig. \ref{rx23_fot}, 
the shape of the orbital modulation does not vary dramatically with wavelength,
allowing us to combine the optical and {\it H} band data in a single set for the
period analysis. Starting from the 0.146\,13~d period value for 2003, we
find a minimum phase dispersion for both optical and infrared data for
the ephemeris:

\begin{equation}
T_{min} (HJD) = 2\,452\,825.837\,6\;(30) + 0.145\,451\,6\;(10)\;E\; .
\label{eq-rx23}
\end{equation}

\subsection{Results from photometry and polarimetry}

\begin{figure}
\includegraphics[width=84mm]{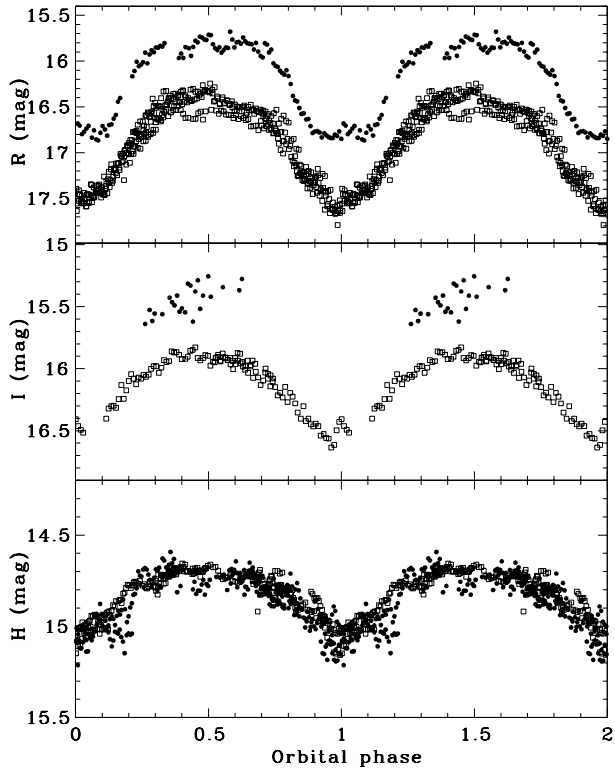}
\caption{Optical and NIR photometry of
\rrxx23: the dots were used for 2003 data and open squares
for 2004 data. The optical and IR data are not simultaneous.}
\label{rx23_fot}
\end{figure}

Figure \ref{rx23_fot}  shows the photometric data folded using the
ephemeris of Equation (\ref{eq-rx23}). The optical magnitudes of 
\rxx23 have
been calculated using USNO~B1.0~0845$-$0644672 as a calibrator ($R2$~=~14.1~mag,
$I2$~=~13.43~mag). The {\it H} band magnitudes are relative to 
2MASS~23161253$-$0529164 (H~=~11.609~$\pm$~0.025~mag) in 2003, and relative 
to 2MASS~23160010$-$0526237 (H~=~12.498~$\pm$~0.025~mag) in 2004.
 
In September, 2003, the $R_C$ band flux of \rxx23 was twice its value in
October, 2004, suggesting a higher accretion state in 2003. In spite of the
poor quality of the $I_C$ band data in 2003, they also indicate a
higher luminosity in 2003. We discarded a flux variation in the main
comparison star by
verifying its flux relative to other field objects. The USNO B1.0 photometry of
the \rxx23 (USNO~B1.0~0845$-$0644683) is $R1$~=~18.550~mag,
$R2$~=~17.190~mag and $I2$~=~17.070~mag.
The {\it H} band flux does not show year-to-year variation, and  the 2MASS
catalogue value ($H$~=~14.798~mag at $\phi_{orb}$ = 0.33) is consistent 
with our data: the long-term variable
component is far less prominent at $1.6 \mu$m. 
However, one should recall that the $H$ band data 
are not simultaneous with the $R_C$ and $I_C$ data. 
The {\it H} band light curve is consistent with the heating of the secondary by the
primary being important in this system, as seen in other CVs.
It can also be seen that
the amplitude of the orbital modulation decreases for longer wavelengths.
 
\begin{figure}
\includegraphics[width=84mm]{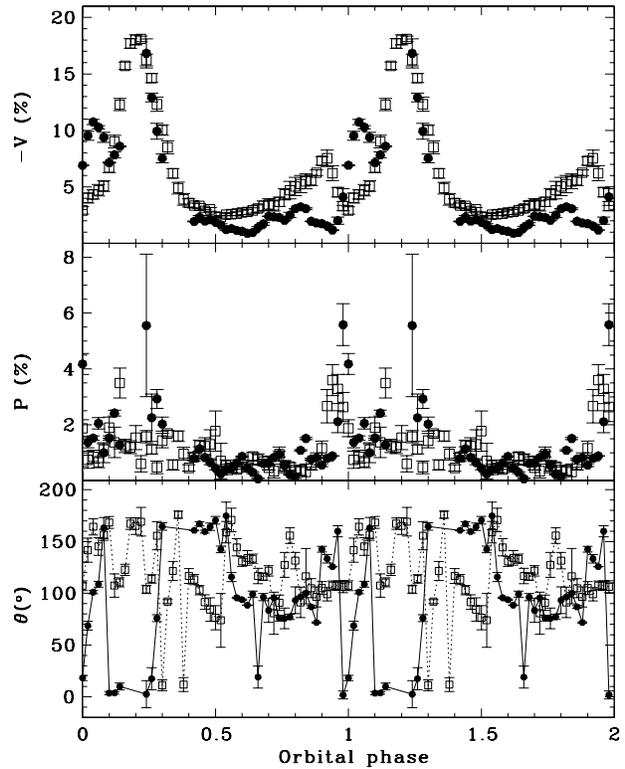}
\caption{Circular and linear polarimetry of
\rxx23 in the $R_C$ band. The data were binned in 50
phase boxes. Dots stands for 2003 data and open squares, for 2004 data.
The circular polarization sign is the instrumental one.}
\label{rx23_pol_r}
\end{figure}

\begin{figure}
\includegraphics[width=84mm]{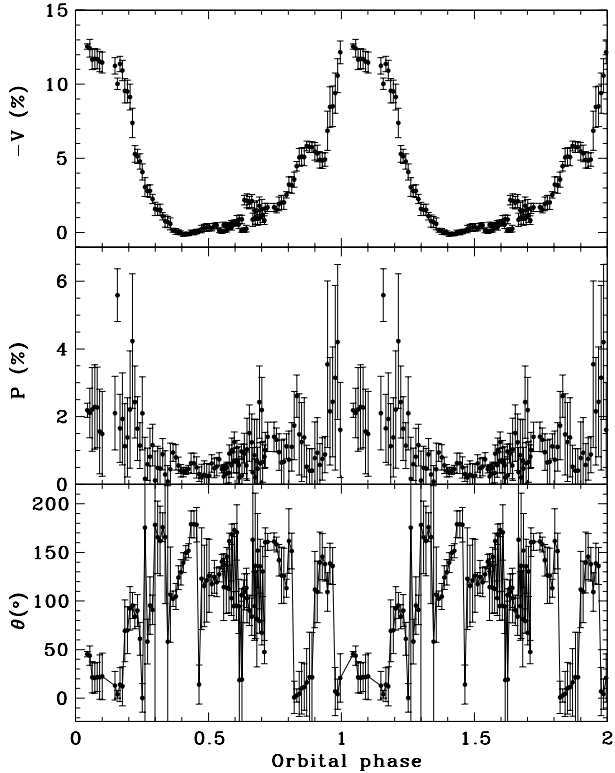}
\caption{Circular and linear polarimetry of
1RXS~J231603.9$-$052713 in the $I_C$ band for 2004 October.
No binning was applied.  The circular polarization sign is the instrumental one.}
\label{rx23_pol_i}
\end{figure}

\begin{figure}
\includegraphics[width=84mm]{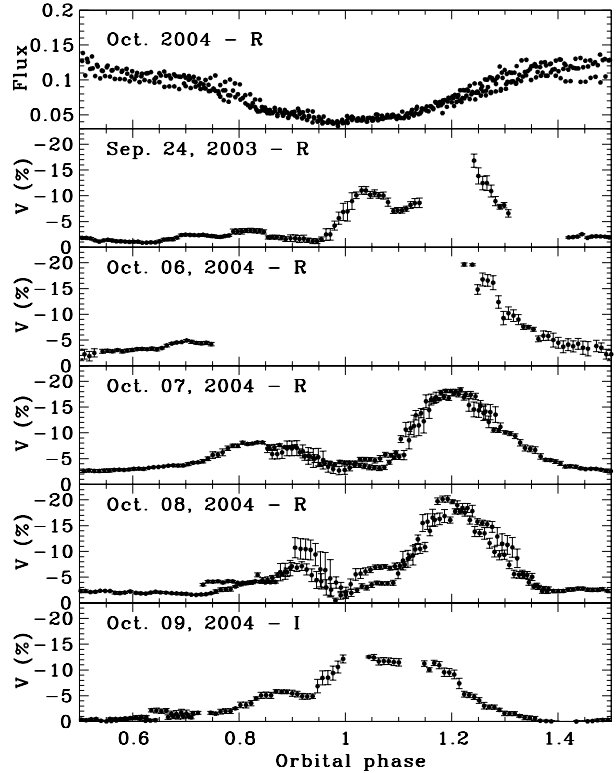}
\caption{Circular polarimetry of
1RXS~J231603.9$-$052713. Each panel represents the data for
one day of observation. The first panel
presents the $R_C$ band flux (in arbitrary units) in 2004 October 06--09.
No binning was applied.  The circular polarization sign is the instrumental one.}
\label{rx23_pol_circ}
\end{figure}

Figures \ref{rx23_pol_r} and \ref{rx23_pol_i} show the circular and 
linear polarimetry of
\rrxx23 in the $R_C$ and $I_C$ bands, respectively. In both
filters, we measured large, strongly variable 
circular polarization, confirming this
object as a polar. In Figure \ref{rx23_pol_r} all the data from
a given year were
binned in 50 orbital phase intervals to provide a better visualization.
The  variability of this object is illustrated in Figure \ref{rx23_pol_circ}
which shows daily circular polarimetry without any binning.

As for 1RXS J161008.0+035222 no change of sign in the circular polarization
is observed. \rxx23 seems, however, to have a more complex accretion
geometry. Therefore the suggestion of a one-pole geometry could be
incorrect.  In a dipolar magnetic geometry, the possibilities are: 
(1) a single pole visible; (2) two poles with the same polarity; or (3) one pole
observed from the top, with the other seen upside down.

The $R_C$ band circular polarization presents a maximum of constant amplitude
near phase 0.2. The interval $ 0.4 < \phi_{orb} < 0.6$  shows small
circular polarization. From phase 0.7 to 1.1 the polarization varies
significantly, presenting variable peaks at different phases in different
occasions.

The phasing of the flux and polarization curves, the asymmetric profile, and the highly
changeable circular polarization makes \rxx23 a very interesting object. 
These properties resemble QQ Vul (Cropper 1998; Schwope et al.
2000b) and may be originated by a complex and variable
accretion geometry.

\subsection{Modelling the optical flux and polarization: a cyclotron model}
\label{cyclotron_model}

Following the same procedure described in the previous section for \rrx16, we tried to fit the
$R_C$
band data
with a simple model of a non-polarized component plus a cyclotron and free-free
emitting region represented by the WM85 models. Considering only the
light-curve, we could identify two families of models: (1) the photometric
minimum corresponds to the time when the accretion column has the smallest
angle with the line of sight and is caused by intrinsic properties of the
accreting region; or (2) the minimum occurs when the column points away 
from the observer
and is caused by occultation of the accreting region by the white-dwarf.

We discard the second family of models for many reasons. First, 
the model fluxes fit the light-curve poorly.
In addition, the accretion 
column in polars tends to be located near the plane perpendicular to
the orbit and connecting the two stars (Liebert \& Stockman 1985). Also,
if the {\it H} band light-curve is due to a heated secondary, the
phasing would be wrong. 

\begin{figure}
\includegraphics[width=84mm]{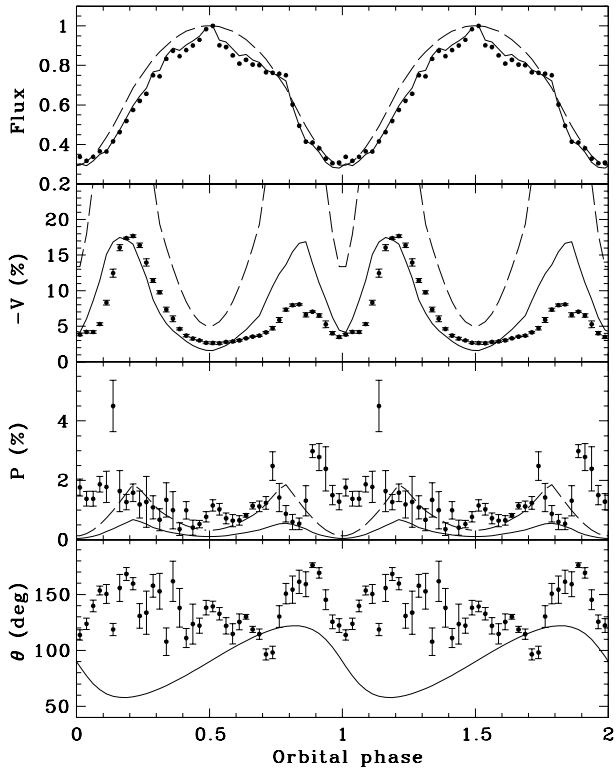}
\caption{Sample fits for
1RXS~J231603.9$-$052713 $R_C$ band data from Oct 07, 2004.
The dashed line represents a two-component model: WM85 plus a constant and
non-polarized emission, in which the WM85 responds   for 73\% of the maximum
flux. The solid line is a three-component  model:
WM85 (24\% at maximum light), a non-polarized constant flux (25\%) 
and a non-polarized modulated
emission (51\%). In both models the parameters of the WM85 component are:
$i~=~40\degr$; $\beta~=~20\degr$;
$|B|$~=~23~MG; T~=~10~keV; $\Lambda$~=~$10^7$.}
\label{rx23_mod_wm85}
\end{figure}

Hereafter, we consider only the first family of models. They provide
inclinations of around 30--50$\degr$, colatitudes
of the magnetic axis ranging from 20 to 30$\degr$ and a magnetic field strength 
of 20--30~MG. These models reproduce very well
the light-curve, but not  the circular polarization curve. In Figure
\ref{rx23_mod_wm85} (dashed line) we show one such a model. It is clear that
the predicted polarization is larger than the
observed: no alternative choice of parameters can improve the fit. A possible
solution is to consider that part of the flux modulation comes from a
non-cyclotron source. To test this hypothesis, we have
constructed a model with a third component having
the same modulation of the observed curve, but with a free scale factor.
We could then reproduce well the light-curve and the
circular polarization peak in phase 0.2. (Figure \ref{rx23_mod_wm85}, solid
line).
In Section \ref{sec_heated_secondary}, we return
to this discussion using a plausible model for the modulated component.

The variation of position angle cannot be described using the standard radial
accretion geometry of Stockman (1977).
Cropper (1989) has calculated the behavior of
the position angle with orbital phase for accretion along the lines of 
a dipolar magnetic field. The departure from the radial model tends to be
larger for small values of the threading radius -
the radius where the magnetic field line crosses the orbital plane -
(see his Figs. 3 and 4). Therefore, the monotonic increase in
position angle near phase 0.0 and 0.25 in 2003 data could be interpreted as
non-radial accretion with a small threading radius ($\la$ 50 radius of the 
white-dwarf). Alternatively, the occurrence of two of these features 
may indicate the presence of two accreting regions. 

\subsection{Modelling the optical and infrared flux: a cyclotron model plus
a heated secondary}
\label{sec_heated_secondary}

As shown in the previous section, there is a suggestion of a modulation
in the $R_C$ band originated by non-cyclotron emission. A potential source 
would be the contribution from the heated secondary star. This hypothesis is
studied below using the code presented in Appendix \ref{lcm}, which includes
(1) the white-dwarf contribution, (2) a heated secondary star, and (3) an
arbitrary component which we have selected from the grid of models from WM85.

A first attempt to fit the $R_C$ and {\it H} band photometry with a simple
model in
which only the secondary star (including heating effect) and white-dwarf
were present failed in the sense that too much $R_C$ flux is
predicted if we fit the $H$ band data. It indicates 
that a third component exists even in the NIR. 
This is not surprising since we
know that there is additional light in optical wavelengths from 
cyclotron emission in this binary.
If this cyclotron emission is also present in the $H$ band,
we have one more constraint to the magnetic field (B) value, since that
emission occurs at wavelengths larger than a given limit
inversely proportional to B. This discards models with large
values of $B$. 

We chose two models (from the first family as defined in
Sec. \ref{cyclotron_model}) from the grid of WM85 that
describe the polarimetry reasonably well, but have distinct
behavior in the near infrared: the first is constant in the $H$ band.
They have the following parameters:

\begin{enumerate}

\item $i~=~40\degr$, $\beta~=~20\degr$, $|B|$~=~23~MG, T~=~10~keV,
$\Lambda$~=~$10^7$ (Figure \ref{rx23_mod_wm85});

\item $i~=~40\degr$, $\beta~=~30\degr$, $|B|$~=~27.1~MG, T~=~10~keV,
$\Lambda$~=~$10^6$.

\end{enumerate}

The analysis below takes into account the simultaneous fit of the 
light-curves in the $R_C$, $I_C$ and $H$ bands. 
We incorporated the cyclotron contribution into our model, fixing
its fraction relative to the total light at $\phi_{orb}=0.5$ to be 0.25
in the $R_C$ band. In the other bands, we fix the cyclotron contribution
in such a way that the proportion relative to the $R_C$ band is that
given by the WM85 models.
The inclination of the system is fixed at
40 degrees, to be consistent with the polarimetric models chosen from the
WM85 tables. The results depend very little on the inclination for the
30-50 degrees range. All modelling assumes the primary to follow the 
Hamada-Salpeter mass-radius relation. The allowed range of primary masses
is 0.4~--~1.1~$M_\odot$. The Roche-lobe filling star is forced to follow the
mass-radius relation of Chabrier \& Baraffe (1997). Reddening is assumed
to be E($B-V$)=0.037  estimated using the
procedure from Schlegel, Finkbeiner \& Davis (1998). Unless otherwise stated,
the following discussion is valid for both polarimetric models.

We would like to note that our photometry is calibrated
against the USNO B1.0 magnitudes of USNO B1.0 0845-0644672 (R2=14.1, 
I2=13.43). The convertion between $R2$ and Landolt's $R_C$ for this object
indicates a difference of only 0.02 mag (Kidger 2003). We assume the $I2$ 
magnitude also to be the same as Landolt's.

\begin{figure}
\hspace{-1.0cm}
\includegraphics[width=100mm]{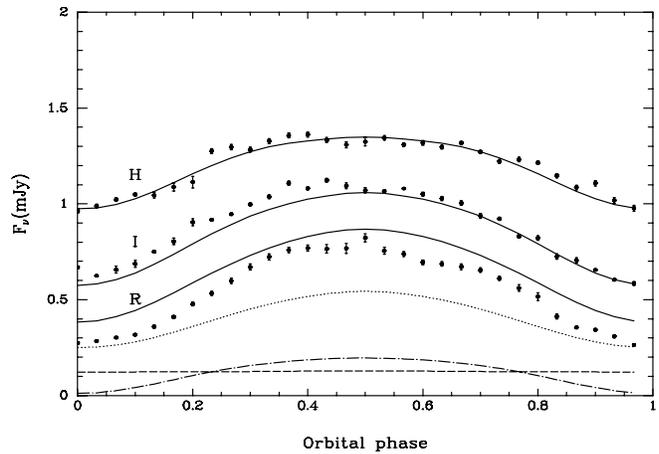}
\caption{Heated secondary models of 1RXS~J231603.9$-$052713
in the $R_C$, $I_C$, and $H$ bands for model $i$. The individual contributions are
shown for the $R_C$ band, with the ellipsoidal+heating component shown as a dotted
line, the white-dwarf contribution shown as a dashed line and the cyclotron
component as dot-dashed (model i). The total flux is shown as a continuum line.}
\label{rx23_modflux_i}
\end{figure}

\begin{figure}
\hspace{-1.0cm}
\includegraphics[width=100mm]{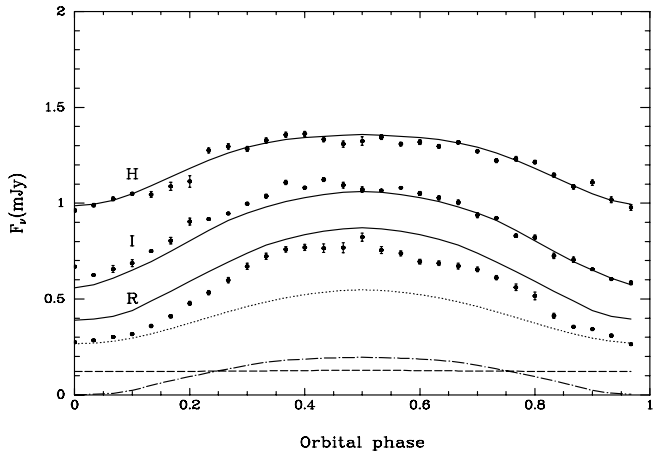}
\caption{Same as in Figure \ref{rx23_modflux_i}, for model ii.}
\label{rx23_modflux_ii}
\end{figure}

Table \ref{tab-mod1} shows the parameters obtained for the best fits for each 
WM85 model which are shown in Figures \ref{rx23_modflux_i} and
\ref{rx23_modflux_ii}
superimposed in the $H$, $I_C$ and $R_C$ band light-curves.
The $H$ and $I_C$ fluxes are well fitted by the models, but
the amplitude of the modulation in the $R_C$ band model is
smaller than the observed.

\begin{table}
\caption{Parameters of the emission models for \rxx23 shown in Figures 
\ref{rx23_modflux_i} and \ref{rx23_modflux_ii}}
\label{tab-mod1}
\begin{tabular}{@{}lll@{}}
\hline
Parameter & Model $i$ & Model $ii$ \\
\hline
$T_1$ & 55\,575 K & 54\,626 K \\
$T_2$ & 3\,280 K & 3\,320 K \\
$R_1/a$ & 0.0052 & 0.0053 \\
$M_1$ & 1.09 $M_\odot$ & 1.09 $M_\odot$ \\
$M_2$ & 0.39 $M_\odot$  & 0.39 $M_\odot$ \\
Distance, d & 416 pc & 418 pc \\
$\chi$ & 15.57 & 15.45 \\
\hline
\end{tabular}
\end{table}

A word about the figure of merit used in the optimization of our models.
We adopted $\chi = \sqrt{\chi_R^2 + \chi_I^2 + \chi_H^2}$ where 
$\chi_X^2 = \sum_{i=1}^N \frac{(O_i - C_i)^2}{\sigma_i^2} $ stands for
$\chi^2$ in the $X$ band. The high values of $\chi$ shown in Table
\ref{tab-mod1}
are due to the fact that the errors $\sigma_i$ in the 30 phase bins take
into account only the statistical uncertainties in the bins themselves.
Flickering, for example, adds additional uncertainties that are not
considered and could only be estimated with longer sets of data.

The temperature of the primary is large when compared with that observed in
cataclysmic variables in general, $\sim 16\,000$ K (Sion 1999) and its presence
could in principle be detected photometrically, specially at $\phi_{orb}=0.0$
where it contributes with a fraction of about 50\% in the $R_C$ band.
However, that high temperature source may be better associated with a
small region in the white-dwarf photosphere reprocessing the 
emission produced in the accretion/shock region.
Such a component is very important in polars and contributes in
soft X-rays and ultraviolet wavelengths.

The temperature of the secondary star is 3\,280~K for  model i. For d~=~416~pc
this corresponds to $M_H$ = +7.5 and spectral type in the range M3-M4 V. The
contribution of the secondary star alone at $\phi_{orb}~=~0.0$ would be 
0.871~mJy corresponding to $H$~$\sim$~15.19. This is in good agreement
with the observed minimum value of $H$~=~15.07 in the binned light-curve.

For model ii, with $T_2$ = 3\,320~K we have an object slightly cooler than M3V
and for d~=~418~pc, we have $M_H$~=~+7.14 at $\phi_{orb}~=~0.0$. Our model predicts
a contribution of 0.903~mJy from the secondary star at $\phi_{orb}~=~0.0$. This
is equivalent to apparent $H$~=~15.13 and agrees well with the observed value.

\begin{figure}
\hspace{-1.0cm}
\includegraphics[width=100mm,clip=true]{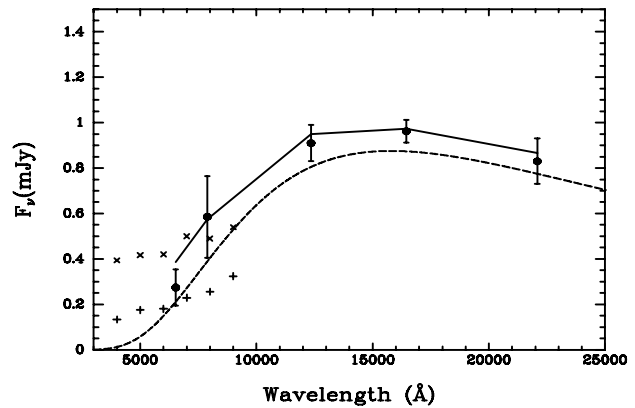}
\caption{The spectral energy distribution for \rxx23 at minimum light (see
text). Model i has been superimposed on the data. The dashed line represents 
a 3\,280 K blackbody curve.}
\label{rx23_sed_i}
\end{figure}

Figure \ref{rx23_sed_i} shows a comparison of the results of
model i with the observed spectral 
energy distribution (SED) at $\phi_{orb}=0.0$. 
The comparison for model ii is 
not shown since it is visually identical to Figure \ref{rx23_sed_i}. 
One can see that the $H$ band measurement has the
smallest uncertainty since our photometry was calibrated against 2MASS
objects. The $J$
and $Ks$ band points are inferred from the 2MASS data.
They were obtained from an interpolation of
the amplitude of the photometric modulation (for the $J$ band) and an
extrapolation for the $K$ band. The model amplitudes are
0.47 mag and 0.20 mag, for the $J$ and $K$ bands, respectively. 
The $R$ and $I$ measurements are from the USNO catalog, therefore, 
non-simultaneous with our $H$ band measurement. 
The uncertainty in the
$R_C$ and $I_C$ measurements comes from the USNO calibration and from the fact that
these measurements are not simultaneous with the NIR data. The crosses and
pluses in Figure \ref{rx23_sed_i}
show the SEDs from measurements in 1991 and 1995
(Thomas et al. 1998 and Thomas, H.-C. private communication). The solid
straight-line segments correspond to the output of our model (at
$\phi_{orb}=0.0$) considering the sum of all components. Notice that at this
phase we expect the smallest contamination from the heating of the secondary
star. The dashed curve shows the contribution of the secondary star if
there was no heating from the white-dwarf.

\begin{figure}
\includegraphics[width=84mm]{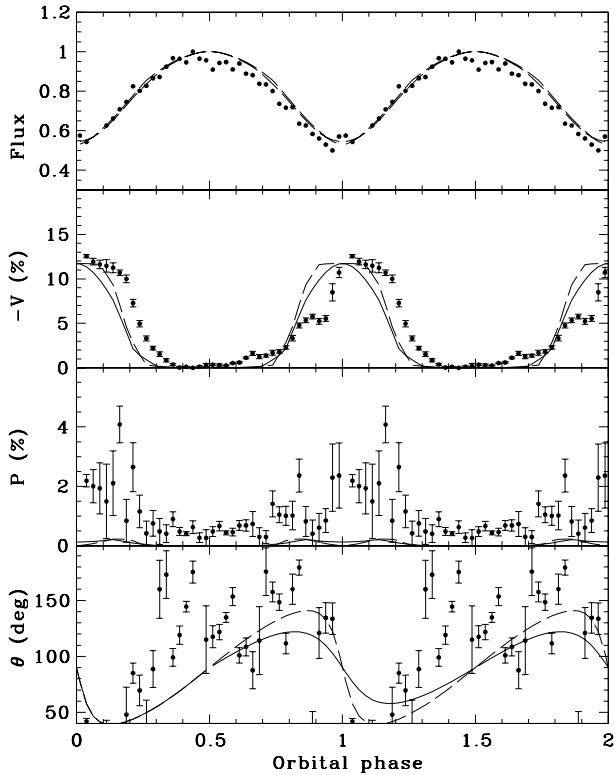}
\caption{Resulting flux and polarization for 1RXS~J231603.9$-$052713 $I_C$ band 
using the heated secondary model. The solid lines represents the model i
results. The dashed line is used to shown the model ii.}
\label{rx23_mod_poli}
\end{figure}

Figure \ref{rx23_mod_poli} shows the results for the $I_C$ band applying
the two best fits for flux and the WM85 models to calculate the polarization. As for
the $R_C$ band, the overall observed level and shape are reproduced,
but the details of the curve needs a more refined modelling.

\section{CONCLUSIONS}

We present polarimetric data for two {\it ROSAT} candidates to 
AM Her systems: \rx16 and \rrxx23. Optical
photometry of \rx16 and optical/NIR photometry of
\rxx23 were obtained. Both systems show high values
of circular polarization confirming them as polars. 
We also presented new ephemerides for the two systems.

A simple model of a point-like accretion region in the 
white-dwarf plus a constant component is able to reproduce the gross features of 
\rrx16. It corresponds to a system seen approximately face-on, 
with a (dipolar) magnetic field aligned with the rotation axis
of the white-dwarf. The magnetic field strength was estimated to be in 
the 10--20~MG range. Our model predicts small values for
the linear polarization which are consistent with the data. 
Another possible model assumes an inclination around
60\degr\ and $\beta \approx 20\degr$. The flux and circular polarization
are well fitted, but the model linear polarization peak is
not present in the data. The averaged
linear polarization of \rx16 is consistent with an interstellar origin,
according to the polarimetry of field stars.

Our $R_C$ band observations of \rxx23 have been taken in two different 
brightness states which show, however, similar light-curve shapes.
The dependence of the $R_C$ band polarization with the orbital
phase shows considerable variation in time-scales of days or years,
but its maximum is apparently constant around 20\%.
The secondary seems to be an important
source of light even in the $R_C$ band. We have presented a model that
reproduces the main features of flux and polarization in
all observed bands. It considers three components:
a white-dwarf, a heated secondary and an accretion column. 
The suggested parameters for the system are:
$T_1 \approx 54\,000 K$ (probably associated with the accretion region);
$T_2 \approx 3\,250 K$;
$R_1/a \approx  0.0052$;
$M_1 \approx 1.10 M_\odot$;
$M_2 \approx 0.39 M_\odot$;
distance $\approx$ 410~pc;
$i~\approx~ 40\degr$; $\beta~=~20$--$30\degr$; $|B|$~=~20--30~MG.
A point-like accretion region is clearly not enough to fit the asymmetry 
of the \rxx23 polarization. An extended and inhomogeneous 
accretion region may better reproduce the data. The position angle of
the linear polarization suggests non-radial accretion in this
system.

\section*{ACKNOWLEDGEMENTS}

We thank R. Downes for clarifying the catalogue entry information on the
orbital period of \rrxx23 and H.-C. Thomas for providing unpublished \rxx23
spectra. We also acknowledge an anonymous referee by his(her)
suggestions. This work was partially supported by Fapesp (CVR: Proc. 2001/12589-1).
This research has made use of: the USNOFS Image and Catalogue Archive
operated by the United States Naval Observatory, Flagstaff Station
(http://www.nofs.navy.mil/data/fchpix/); the SIMBAD database,
operated at CDS, Strasbourg, France;  and the
NASA's Astrophysics Data System Service.

\appendix

\section{A model for optical and infrared flux in CVs: white-dwarf plus a heated
secondary}
\label{lcm}

\subsection{Light-curve modelling}

Our code is similar to the one described in Zhang, Robinson \& Nather (1986).
The secondary star always fills its Roche lobe and the primary star is
spherical. Both are treated as blackbodies.

The stars' surface are represented by a mesh
of 100 $\times$ 100 elements of area.
Given a mass ratio, for each point of the grid corresponding to the secondary
star, we calculate the parameters of the Roche equipotential surface. 
This allows us to compute the
local gravity and the gravity darkening via $T=T_{pole} (g/g_{pole})^{\beta}$.
Here $T_{pole}$ is the temperature of the far pole of the secondary star.
We follow Lucy (1967) and use $\beta=0.08$ for a typical secondary in a CV.

The effect of the irradiation of the primary star on the secondary is
calculated with a simple approximation: $T^4 = T^4_0 + (1-a) F_{irr} / \sigma$.
Here $T_0$ is the grid-point temperature without irradiation,
$a$ is the albedo of the secondary star, $F_{irr}$ is the irradiation flux
on the secondary and $\sigma$ is the Stefan-Boltzmann constant. All points
on the secondary visible from the spherical primary are affected by
irradiation. Both the distance and projected areas of 
the respective surface grid elements are taken into account in the 
calculations. We use $a=0.5$ in our simulations, according to what is
expected for convective stars (Rucinski 1969). 

The limb-darkening is assumed to be linear, that is, 
$I(\mu)/I_0 = 1 - u(1-\mu)$, where $\mu$ is the cosine of the angle from the
center to the border of the star, and 
$u$ is the limb-darkening coefficient. Considering the range of wavelengths 
examined 
in our modelling we adopted $u = 0.6$ for the secondary (Claret 1998). We 
followed 
Thorstensen \& Armstrong (2005) and explored a range of values for the 
limb-darkening 
of the white dwarf. The results depend very little on the particular value 
and we chose $u=0.7$ for the primary.

Since the orbital period and masses of the components define the size of the
binary, absolute dimensions of the grid elements can be used in the
calculation of the flux in the direction of the observer. This means that
the distance can be used as a parameter to be determined from the data.
For a low-inclination system for which the secondary temperature is zero
and a non limb-darkened  spherical primary of 1R$_{\odot}$ with a temperature
of 5000 K at a distance of 1 kpc, the model correctly predicts a flux of 2.05
mJy in the $V$ band.
 
The output of the light-curve synthesis program is the sum of the flux
components due to the secondary star (ellipsoidal variation plus heating of
the secondary), the contribution of the white-dwarf and an optional third
component (e.g., due to cyclotron emission) whose fraction is a parameter 
that can be fixed or determined from the fit to 
the data, according to the information available. 

The main advantage of a simplified light-curve synthesis code like ours is
speed of execution. The fitted parameters are:

\begin{enumerate}

\item orbital inclination;

\item temperature of the primary;

\item pole temperature of the secondary;

\item masses of the components;

\item distance;

\item fraction of total light at $\phi_{orb}=0.5$ due to cyclotron emission.

\end{enumerate}

The radius of the primary is forced to follow the Hamada-Salpeter relation
for Helium white-dwarfs. The range of masses was limited to 
$0.4~M_\odot~<~M_1~<~1.1~M_{\odot}$
according to the observed distribution of field white-dwarfs (Provencal et al.
1998)

The mass of the secondary was allowed to be in the range 
$0.08~M_\odot~<~M_2~<~1.0~M_\odot$. We forced the secondary star to follow the 
mass-radius relation of Chabrier \& Baraffe (1997).
 
\subsection{Monte Carlo Markov Chain}

The approach of using Monte Carlo Markov Chains (MCMC) for the estimation
of parameters has the advantage of a good flexibility on deciding which
parameters
are to be estimated or kept fixed and gives very robust intervals of confidence
for the parameters. A review on MCMC can be found in Gilks, Richardson \&
Spiegelhalter (1996).

We used the MCMC to sample the {\it a posteriori} distribution of the
parameters $\Theta$ of our model as given by the Bayes theorem:

$$ p(\Theta | D) \propto p(D|\Theta)p(\Theta) , $$

\noindent where $p(D|\Theta)$ is the usual likelyhood of the data
given the parameters $\Theta$
and $p(\Theta)$ is any prior information on the parameters. For example, if 
a system does not show eclipses, $p(\Theta)$ may be considered null for
inclination 
angles larger than $\approx 70 \deg$.

The marginalized distributions of the parameters of long Markov Chain allow
us to derive their localization (e.g., via the mode of the distribution) 
and confidence intervals by simple integration along the distribution.  

\end{document}